# Quadratic Matrix Inequality Approach to Robust Adaptive Beamforming for General-Rank Signal Model

Yongwei Huang, *Senior Member, IEEE*, Sergiy A. Vorobyov, *Fellow, IEEE*, and Zhi-Quan Luo, *Fellow, IEEE*

***Abstract*—The worst-case robust adaptive beamforming problem for general-rank signal model is considered. This is a nonconvex problem, and an approximate version of it (obtained by introducing a matrix decomposition on the presumed covariance matrix of the desired signal) has been well studied in the literature. Different from the existing literature, herein however the original beamforming problem is tackled. Resorting to the strong duality of linear conic programming, the robust adaptive beamforming problem for general-rank signal model is reformulated into an equivalent quadratic matrix inequality (QMI) problem. By employing a linear matrix inequality (LMI) relaxation technique, the QMI problem is turned into a convex semidefinite programming problem. Using the fact that there is often a positive gap between the QMI problem and its LMI relaxation, an approximate algorithm is proposed to solve the robust adaptive beamforming in the QMI form. Besides, several sufficient optimality conditions for the nonconvex QMI problem are built. To validate our results, simulation examples are presented, which also demonstrate the improved performance of the new robust beamformer in terms of the output signal-to-interference-plus-noise ratio.**

## I. Introduction

Robust adaptive beamforming techniques provide a powerful approach to significantly improve the array output signal-to-interference-plus-noise ratio (SINR) and other performance metrices such as mainlobe width and sidelobe levels. Here the robustness typically means the ability of a method to perform well under an imperfect knowledge about the source, propagation and sensor array. In particular, when it is difficult to obtain the knowledge of the desired signal covariance matrix, a mismatch between the presumed and actual source covariance matrices causes dramatic performance degradation, and the robust adaptive beamforming techniques are very efficient against the mismatch [1], [2].

Y. Huang is with School of Information Engineering, Guangdong University of Technology, University Town, Guangzhou, Guangdong, 510006, China. Email: ywhuang@gdut.edu.cn.

S. A. Vorobyov is with Department of Signal Processing and Acoustics, School of Electrical Engineering, Aalto University, Konemiehentie 2, 02150 Espoo, Finland. Email: svor@ieee.org.

Z.-Q. Luo is with School of Science and Engineering, the Chinese University of Hong Kong (Shenzhen), Longgang, Shenzhen, Guangdong 518172, China. Email: luozq@cuhk.edu.cn.

Many robust adaptive beamforming approaches have been proposed for the scenario of a rank-one signal model (see [3] and reference therein). However, it is of practical interest to consider a general-rank signal model, as the signal source often can be incoherently scattered, and robust adaptive beamforming for general-rank signal models becomes necessary.

In [4], an efficient robust adaptive beamforming technique for general-rank model has been proposed, and a closed-form beamformer has been computed in terms of a principal eigenvector of the product between inverse of the sample data covariance matrix and the worst-case of the presumed covariance for the signal.

The authors of [5] have presented a new method to the robust adaptive beamforming with general-rank signal models, taking into account a positive semidefinite (PSD) constraint over the mismatched signal covariance matrix (the presumed covariance plus errors). The resultant robust beamforming problem has been formulated by introducing a matrix decomposition (e.g. spectral or Cholesky type) of the presumed signal covariance matrix and putting the error term into both of the matrices obtained from the decomposition. It is different from considering the worst-case of the mismatch signal covariance matrix. It turns out that the robust problem is a nonconvex quadratic program, and an iterative algorithm using semidefinite programming (SDP) has been proposed to find a suboptimal solution.

In [6], two beamformers have been derived in closed-form for the robust adaptive beamforming problem established in [5], leading to low complexity robust beamformers. Under the assumption that the interference is well separated from the signal, the authors of [7] have proposed a method based on SDP relaxation and bisection search to solve the robust beamforming problem formulated in [5].

In [8], [9], the aforementioned beamforming problem is termed as a difference-of-convex (DC) optimization problem, and a polynomial time DC (POTDC) algorithm has been proposed. The authors show that the POTDC converges to a local optimal solution, and under the condition that the error norm bound is sufficiently small, the local solution is indeed a globally optimal solution.

Finally, a more computationally efficient approach to the aforementioned robust adaptive beamforming problem via matrix decomposition of the presumed signal covariance matrix and putting the error term into the matrices obtained after decomposition has been recently developed in [10]. This approach is based on sequential inner second-order cone programming (SOCP) approximations of the problem, and helps to significantly reduce the order of the computational complexity, but yet does not address the original problem.

In this paper,[1] we address the original robust adaptive beamforming problem for general-rank signal model, i.e., address the problem without performing a matrix decomposition over the presumed signal matrix (like what has been done in [5]). Resorting to the strong duality theorem of SDP (see e.g. [12]), the robust problem is reformulated into a nonconvex quadratic matrix inequality (QMI) problem. To tackle the problem, we employ a linear matrix inequality (LMI) relaxation and turn the QMI problem into a convex LMI problem. Based on an optimal solution of the LMI problem, we propose an approximate algorithm to find a solution for the robust adaptive beamforming problem. Besides, sufficient optimality conditions for the QMI problem are established that guarantee a globally optimal solution for it. Finally, we pay a revisit to the dual problem of the SDP relaxation of the robust adaptive beamforming problem for general-rank signal model and conclude that the SDP relaxation is tighter than the conventional minimax relaxation for the problem in a maximin form.

The paper is organized as follows. In Section II, we introduce the signal model and formulate the robust adaptive beamforming problem for general-rank signal model. In Section III, we build a problem reformulation to QMI problem, and propose a deterministic LMI relaxation-based approximate algorithm for it. Sufficient conditions for the tightness of the LMI relaxation are built in Section IV. In Section V, we pay a revisit to the SDP relaxation for the robust beamforming problem. Section VI includes illustrative numerical examples and Section VII draws some concluding remarks.

*Notation*: We adopt the notation of using boldface for vectors $\boldsymbol{a}$ (lower case), and matrices $\boldsymbol{A}$ (upper case). The transpose operator and the conjugate transpose operator are denoted by the symbols $(\cdot)^T$ and $(\cdot)^H$ respectively. The notation tr$(\cdot)$ stands for the trace of the square matrix argument; $\boldsymbol{I}$ and $\boldsymbol{0}$ denote respectively the identity matrix and the matrix (or the row vector or the column vector) with zero entries (their size is determined from the context). The letter $j$ represents the imaginary unit (i.e. $j = \sqrt{-1}$), while the letter $i$ often serves as index in this paper. For any complex number $x$, we use $\Re(x)$ and $\Im(x)$ to denote respectively the real and the imaginary parts of $x$, $|x|$ and arg$(x)$ represent the modulus and the argument of $x$, and $x^*$ ($\boldsymbol{x}^*$ or $\boldsymbol{X}^*$) stands for the (component-wise) conjugate of $x$ ($\boldsymbol{x}$ or $\boldsymbol{X}$). The Euclidean norm (the Frobenius norm) of the vector $\boldsymbol{x}$ (the matrix $\boldsymbol{X}$) is denoted by $\|\boldsymbol{x}\|$ ($\|\boldsymbol{X}\|$). The curled inequality symbol $\succeq$ (and its strict form $\succ$) is used to denote generalized inequality: $\boldsymbol{A} \succeq \boldsymbol{B}$ means that $\boldsymbol{A}-\boldsymbol{B}$ is an Hermitian positive semidefinite matrix ($\boldsymbol{A} \succ \boldsymbol{B}$ for positive definiteness). The space of Hermitian $N \times N$ matrices (the space of real-valued symmetric $N \times N$ matrices) is denoted by $\mathcal{H}^N$ ($\mathcal{S}^N$), and the set of all positive semidefinite matrices in $\mathcal{H}^N$ ($\mathcal{S}^N$) is denoted by $\mathcal{H}_+^N$ ($\mathcal{S}_+^N$), while Rank $(\cdot)$ and $\mathsf{E}[\cdot]$ represent the rank of a matrix argument and statistical expectation, respectively. The notations $\lambda_{\max}(\boldsymbol{X})$ ($\lambda_{\min}(\boldsymbol{X})$) and $\boldsymbol{X}^{1/2}$ stand respectively for the maximal (minimal) eigenvalue of the square matrix $\boldsymbol{X}$ and the square root matrix if $\boldsymbol{X} \succeq \boldsymbol{0}$, while $\lambda_i(\cdot)$ stands just for $i$th eigenvalue of a square matrix argument. Finally, $v^\star(\cdot)$ represents the optimal value of problem $(\cdot)$.

[1] Some preliminary results have been also reported in [11].

## II. Signal Model and Problem Formulation

The output signal of a narrowband receive beamformer can be written as

$$x(t) = \boldsymbol{w}^H \boldsymbol{y}(t)$$

where $\boldsymbol{w}$ is the $N \times 1$ vector of beamformer complex weight coefficients, $\boldsymbol{y}(t)$ is the $N \times 1$ complex snapshot vector of array observations, and $N$ is the number of antenna elements of the array. The observation vector is given by

$$\boldsymbol{y}(t) = \boldsymbol{s}(t) + \boldsymbol{i}(t) + \boldsymbol{n}(t) \tag{1}$$

where $\boldsymbol{s}(t)$, $\boldsymbol{i}(t)$, and $\boldsymbol{n}(t)$ are the statistically independent components of the desired signal, interference, and noise, respectively. The output SINR of the beamformer is given by

$$\text{SINR} = \frac{\boldsymbol{w}^H \boldsymbol{R}_s \boldsymbol{w}}{\boldsymbol{w}^H \boldsymbol{R}_{i+n} \boldsymbol{w}} \tag{2}$$

where the desired signal covariance matrix is $\boldsymbol{R}_s \triangleq \mathsf{E}[\boldsymbol{s}(t)\boldsymbol{s}^H(t)]$ and the interference-plus-noise covariance matrix is $\boldsymbol{R}_{i+n} \triangleq \mathsf{E}[(\boldsymbol{i}(t)+\boldsymbol{n}(t))(\boldsymbol{i}(t)+\boldsymbol{n}(t))^H]$. Note that the SINR value (2) is unaltered when the norm of beamvector $\boldsymbol{w}$ changes or a phase rotation is performed over $\boldsymbol{w}$. Matrix $\boldsymbol{R}_s$ herein can be of rank one or higher, i.e., Rank $(\boldsymbol{R}_s) \in \{1, \ldots, N\}$. Both rank-one (corresponding to the case of the point source) and higher-rank $\boldsymbol{R}_s$ are common in many practical situations occurring in wireless communications, radar and sonar (see [1], [3]–[5]).

Suppose that $\boldsymbol{R}_s$ and $\boldsymbol{R}_{i+n}$ are known perfectly in some ways, then an optimal beamforming problem of maximizing the SINR can be cast into:

$$\underset{\boldsymbol{w} \neq \boldsymbol{0}}{\text{maximize}} \quad \frac{\boldsymbol{w}^H \boldsymbol{R}_s \boldsymbol{w}}{\boldsymbol{w}^H \boldsymbol{R}_{i+n} \boldsymbol{w}} \quad . \tag{3}$$

It is evident that (3) is equivalent to the following quadratically constrained quadratic programming (QCQP) problem:

$$\begin{array}{ll} \underset{\boldsymbol{w}}{\text{maximize}} & \boldsymbol{w}^H \boldsymbol{R}_s \boldsymbol{w} \\ \text{subject to} & \boldsymbol{w}^H \boldsymbol{R}_{i+n} \boldsymbol{w} = 1, \end{array} \tag{4}$$

and that the optimal value for (4) and thus also (3) is $\lambda_{\max}(\boldsymbol{R}_{i+n}^{-1/2} \boldsymbol{R}_s \boldsymbol{R}_{i+n}^{-1/2})$ (assuming that $\boldsymbol{R}_{i+n}$ is of full rank), and the optimal solution is a principal eigenvector of $\boldsymbol{R}_{i+n}^{-1/2} \boldsymbol{R}_s \boldsymbol{R}_{i+n}^{-1/2}$ (an eigenvector corresponding to the largest eigenvalue).

In practical applications, however, the interference-plus-noise covariance matrix $\boldsymbol{R}_{i+n}$ is not available. Thus, the sample covariance matrix for $\boldsymbol{R} = \mathsf{E}[\boldsymbol{y}(t)\boldsymbol{y}^H(t)]$, that is,

$$\hat{\boldsymbol{R}} = \frac{1}{T} \sum_{t=1}^{T} \boldsymbol{y}(t) \boldsymbol{y}^H(t) \tag{5}$$

is used to replace $\boldsymbol{R}_{i+n}$ in the optimal beamforming design (3). In (5), $T$ stands for the number of training snapshots. On the other hand, often the signal covariance matrix $\boldsymbol{R}_s$ is only known imperfectly; in other words, there is always a certain mismatch between the presumed signal covariance matrix $\hat{\boldsymbol{R}}_s$ and the actual signal covariance matrix. The beamvector obtained by maximizing the SINR defined by $\hat{\boldsymbol{R}}_s$ and $\hat{\boldsymbol{R}}$ (without taking into account the error terms), however, leads to performance degradation of the array. Therefore, in order to improve the performance, robust adaptive beamforming has been considered, and there are a number of papers on this subject (for example, see [2], [3] for an overview and references therein) in the last two decades.

Herein, let us consider robust adaptive beamforming with general-rank $\boldsymbol{R}_s$ (we refer to [13] for rank one $\boldsymbol{R}_s$), aiming to develop a new efficient method to find a robust beamformer with improved performance. Toward the end, the following robust adaptive beamforming problem maximizing the worst-case SINR is studied:

$$\underset{\boldsymbol{w}\neq\boldsymbol{0}}{\text{maximize}} \quad \underset{\boldsymbol{\Delta}_1\in\mathcal{B}_1,\boldsymbol{\Delta}_2\in\mathcal{B}_2}{\text{minimize}} \quad \frac{\boldsymbol{w}^H\left(\hat{\boldsymbol{R}}_s+\boldsymbol{\Delta}_2\right)\boldsymbol{w}}{\boldsymbol{w}^H\left(\hat{\boldsymbol{R}}+\boldsymbol{\Delta}_1\right)\boldsymbol{w}} \tag{6}$$

where the uncertainty sets $\mathcal{B}_1$ and $\mathcal{B}_2$ are given by

$$\mathcal{B}_1=\left\{\boldsymbol{\Delta}_1\in\mathbb{C}^{N\times N} \mid \|\boldsymbol{\Delta}_1\|\leq\gamma,\ \hat{\boldsymbol{R}}+\boldsymbol{\Delta}_1\succeq\boldsymbol{0}\right\}, \tag{7}$$

and

$$\mathcal{B}_2=\left\{\boldsymbol{\Delta}_2\in\mathbb{C}^{N\times N} \mid \|\boldsymbol{\Delta}_2\|\leq\epsilon,\ \hat{\boldsymbol{R}}_s+\boldsymbol{\Delta}_2\succeq\boldsymbol{0}\right\}, \tag{8}$$

respectively.

Since $\boldsymbol{\Delta}_1$ and $\boldsymbol{\Delta}_2$ are separable, (6) can be recast into:

$$\underset{\boldsymbol{w}\neq\boldsymbol{0}}{\text{maximize}} \quad \frac{\underset{\boldsymbol{\Delta}_2\in\mathcal{B}_2}{\text{minimize}}\ \boldsymbol{w}^H\left(\hat{\boldsymbol{R}}_s+\boldsymbol{\Delta}_2\right)\boldsymbol{w}}{\underset{\boldsymbol{\Delta}_1\in\mathcal{B}_1}{\text{maximize}}\ \boldsymbol{w}^H\left(\hat{\boldsymbol{R}}+\boldsymbol{\Delta}_1\right)\boldsymbol{w}}. \tag{9}$$

Observe that for $\boldsymbol{\Delta}_1\in\mathcal{B}_1$, it follows that

$$\boldsymbol{w}^H\boldsymbol{\Delta}_1\boldsymbol{w}=\text{tr}\left(\boldsymbol{\Delta}_1\boldsymbol{w}\boldsymbol{w}^H\right)\leq\|\boldsymbol{\Delta}_1\|\cdot\|\boldsymbol{w}\boldsymbol{w}^H\|\leq\gamma\|\boldsymbol{w}\|^2$$

and the equalities hold when $\boldsymbol{\Delta}_1=\gamma\boldsymbol{w}\boldsymbol{w}^H/\|\boldsymbol{w}\|^2\in\mathcal{B}_1$. Therefore, the denominator of the objective function of (9) equals

$$\boldsymbol{w}^H\left(\hat{\boldsymbol{R}}+\gamma\boldsymbol{I}\right)\boldsymbol{w}, \tag{10}$$

where $\hat{\boldsymbol{R}}+\gamma\boldsymbol{I}$ represents the worst-case covariance matrix, and $\gamma$ stands for a diagonal loading factor. Thus, (9) can be reexpressed as:

$$\underset{\boldsymbol{w}}{\text{maximize}} \quad \frac{\underset{\boldsymbol{\Delta}_2\in\mathcal{B}_2}{\text{minimize}}\ \boldsymbol{w}^H\left(\hat{\boldsymbol{R}}_s+\boldsymbol{\Delta}_2\right)\boldsymbol{w}}{\boldsymbol{w}^H\left(\hat{\boldsymbol{R}}+\gamma\boldsymbol{I}\right)\boldsymbol{w}}, \tag{11}$$

which is equivalent to the following problem:

$$\begin{array}{ll}\underset{\boldsymbol{w}}{\text{maximize}} & \underset{\boldsymbol{\Delta}_2\in\mathcal{B}_2}{\text{minimize}}\ \boldsymbol{w}^H\left(\hat{\boldsymbol{R}}_s+\boldsymbol{\Delta}_2\right)\boldsymbol{w}\\ \text{subject to} & \boldsymbol{w}^H\left(\hat{\boldsymbol{R}}+\gamma\boldsymbol{I}\right)\boldsymbol{w}=1.\end{array} \tag{12}$$

Regarding the equivalence between problems (11) and (12), the following proposition holds true.

**Proposition II.1** *Problem* (11) *and* (12) *are equivalent to each other, in the sense that they share the same optimal value and if* $\boldsymbol{w}^\star$ *solves* (12)*, then it is optimal for* (11) *too.*

See Appendix A for the proof.

Therefore, we only need to focus on solving the latter maximin problem (12) in order to solve robust adaptive beamforming problem (6). Before proceeding, we present a brief introduction on how the existing works deal with the PSD constraint in feasible set $\mathcal{B}_2$. To cope with it, the objective of (12) is replaced with

$$\boldsymbol{w}^H(\boldsymbol{Q}+\boldsymbol{\Delta}_3)^H(\boldsymbol{Q}+\boldsymbol{\Delta}_3)\boldsymbol{w}, \tag{13}$$

where $\hat{\boldsymbol{R}}_s=\boldsymbol{Q}^H\boldsymbol{Q}$, $\boldsymbol{Q}\in\mathbb{C}^{M\times N}$, $N\geq M=\text{Rank}\,(\boldsymbol{R}_s)$, and the norm of distortion $\boldsymbol{\Delta}_3$ is simply bounded by $\eta$:

$$\mathcal{B}_3=\left\{\boldsymbol{\Delta}_3\in\mathbb{C}^{M\times N} \mid \|\boldsymbol{\Delta}_3\|\leq\eta\right\}. \tag{14}$$

In (13), the covariance matrix has the structure

$$(\boldsymbol{Q}+\boldsymbol{\Delta}_3)^H(\boldsymbol{Q}+\boldsymbol{\Delta}_3)=\hat{\boldsymbol{R}}_s+\boldsymbol{Q}^H\boldsymbol{\Delta}_3+\boldsymbol{\Delta}_3^H\boldsymbol{Q}+\boldsymbol{\Delta}_3^H\boldsymbol{\Delta}_3,$$

which is different from the general form $\hat{\boldsymbol{R}}_s+\boldsymbol{\Delta}_2$. Therefore, the new objective function (13) is a compromise and approximate version for the general form.

With new objective (13), problem (12) is reformulated into:

$$\begin{array}{ll}\underset{\boldsymbol{w}}{\text{maximize}} & \|\boldsymbol{Q}\boldsymbol{w}\|-\eta\|\boldsymbol{w}\|\\ \text{subject to} & \boldsymbol{w}^H\hat{\boldsymbol{R}}\boldsymbol{w}+\gamma\|\boldsymbol{w}\|^2\leq 1,\end{array} \tag{15}$$

following the fact that

$$\min_{\boldsymbol{\Delta}_3\in\mathcal{B}_3}\ \boldsymbol{w}^H(\boldsymbol{Q}+\boldsymbol{\Delta}_3)^H(\boldsymbol{Q}+\boldsymbol{\Delta}_3)\boldsymbol{w} = (\max\{\|\boldsymbol{Q}\boldsymbol{w}\|-\eta\|\boldsymbol{w}\|,0\})^2$$

(see e.g. [9], [14]). It is evident that problem (14) amounts to the following nonconvex problem (cf. [10]):

$$\begin{array}{ll}\underset{\boldsymbol{w}}{\text{minimize}} & \boldsymbol{w}^H\hat{\boldsymbol{R}}\boldsymbol{w}+\gamma\|\boldsymbol{w}\|^2\\ \text{subject to} & \|\boldsymbol{Q}\boldsymbol{w}\|-\eta\|\boldsymbol{w}\|\geq 1.\end{array} \tag{16}$$

which has been addressed in [4]–[10] using different convex approximations, and the global optimality for (16) has not been established generally. In this paper, we, however, focus on solving original robust adaptive beamforming problem (12) (or (11)) without introducing a new objective function like (13).

## III. A Quadratic Matrix Inequality Method to Solve The Robust Adaptive Beamforming Problem For General-Rank Signal Model

In this section, we develop a QMI-based method to solve (12), i.e., robust adaptive beamforming problem for general-rank signal model (6).

To start, the following lemma about the Lagrangian dual problem of a linear conic program is useful.

**Lemma III.1** *The dual problem of*

$$\begin{array}{ll}\underset{\boldsymbol{\Delta}}{\text{minimize}} & \text{tr}\,(\boldsymbol{A}\boldsymbol{\Delta})\\ \text{subject to} & \boldsymbol{R}+\boldsymbol{B}^H\boldsymbol{\Delta}\boldsymbol{B}\succeq\boldsymbol{0}\\ & \|\boldsymbol{C}^H\boldsymbol{\Delta}\boldsymbol{C}\|\leq\epsilon,\end{array}$$

*is the SDP problem:*

$$\begin{array}{ll} \underset{\boldsymbol{X},\boldsymbol{Y}}{\text{maximize}} & -\text{tr}\,(\boldsymbol{R}\boldsymbol{X}) - \epsilon\|\boldsymbol{Y}\| \\ \text{subject to} & \boldsymbol{A} = \boldsymbol{B}\boldsymbol{X}\boldsymbol{B}^H + \boldsymbol{C}\boldsymbol{Y}\boldsymbol{C}^H \\ & \boldsymbol{X} \succeq \boldsymbol{0},\ \boldsymbol{Y} \in \mathcal{H}^N, \end{array}$$

*where* $\boldsymbol{A}$*,* $\boldsymbol{R} \in \mathcal{H}^N$*, and* $\boldsymbol{B}$*,* $\boldsymbol{C} \in \mathbb{C}^{N\times N}$*.*

The proof is straightforward since it can be done in a standard way (see e.g. the derivations between (17) and (24) in [15]), and thus we omit it. In particular, when $\boldsymbol{B} = \boldsymbol{C} = \boldsymbol{I}$, we obtain that

$$\begin{array}{ll} \underset{\boldsymbol{\Delta}}{\text{minimize}} & \text{tr}\,(\boldsymbol{A}\boldsymbol{\Delta}) \\ \text{subject to} & \boldsymbol{R} + \boldsymbol{\Delta} \succeq \boldsymbol{0} \\ & \|\boldsymbol{\Delta}\| \le \epsilon, \end{array}$$

has the dual

$$\begin{array}{ll} \underset{\boldsymbol{X}}{\text{maximize}} & -\text{tr}\,(\boldsymbol{R}\boldsymbol{X}) - \epsilon\|\boldsymbol{X} - \boldsymbol{A}\| \\ \text{subject to} & \boldsymbol{X} \succeq \boldsymbol{0}. \end{array}$$

Clearly, the above primal and dual problems are strictly feasible, assuming that $\boldsymbol{R}$ is PSD (for example, $\epsilon\boldsymbol{I}/(2\sqrt{N})$ is a strictly feasible point for the primal problem, and any positive definite matrix is strictly feasible for the dual problem). Hence the strong duality (see, e.g., [12]) between them holds, which means that they are solvable[2] and their optimal values are equal to each other.

Therefore, it follows that (12) can be reformulated into:

$$\begin{array}{ll} \underset{\boldsymbol{w},\boldsymbol{X}}{\text{maximize}} & \text{tr}\left(\hat{\boldsymbol{R}}_s\left(\boldsymbol{w}\boldsymbol{w}^H - \boldsymbol{X}\right)\right) - \epsilon\left\|\boldsymbol{w}\boldsymbol{w}^H - \boldsymbol{X}\right\| \\ \text{subject to} & \boldsymbol{w}^H\left(\hat{\boldsymbol{R}} + \gamma\boldsymbol{I}\right)\boldsymbol{w} = 1 \\ & \boldsymbol{X} \succeq \boldsymbol{0}, \end{array} \tag{17}$$

and after setting $\boldsymbol{Y} \triangleq \boldsymbol{w}\boldsymbol{w}^H - \boldsymbol{X}$, one can obtain the following problem equivalent to (12):

$$\begin{array}{ll} \underset{\boldsymbol{w},\boldsymbol{Y}}{\text{maximize}} & \text{tr}\,(\hat{\boldsymbol{R}}_s\boldsymbol{Y}) - \epsilon\|\boldsymbol{Y}\| \\ \text{subject to} & \boldsymbol{w}^H\left(\hat{\boldsymbol{R}} + \gamma\boldsymbol{I}\right)\boldsymbol{w} = 1 \\ & \boldsymbol{w}\boldsymbol{w}^H - \boldsymbol{Y} \succeq \boldsymbol{0}. \end{array} \tag{18}$$

The second constraint in problem (18) is a nonconvex QMI constraint. Here we follow the definition that if the terms of the optimization variables in the matrix inequality are quadratic or linear, then the corresponding inequality is called QMI, which is in line with LMI, where all terms of the optimization variables in the matrix inequality must be linear.

There are no existing general approaches in the literature for solving a QMI problem. Herein, we develop an approach for solving particular QMI problem (18). In order to tackle problem (18), we consider its LMI relaxation problem:

$$\begin{array}{ll} \underset{\boldsymbol{W},\boldsymbol{Y}}{\text{maximize}} & \text{tr}\,(\hat{\boldsymbol{R}}_s\boldsymbol{Y}) - \epsilon\|\boldsymbol{Y}\| \\ \text{subject to} & \text{tr}\left((\hat{\boldsymbol{R}} + \gamma\boldsymbol{I})\boldsymbol{W}\right) = 1 \\ & \boldsymbol{W} - \boldsymbol{Y} \succeq \boldsymbol{0},\ \boldsymbol{W} \succeq \boldsymbol{0}. \end{array} \tag{19}$$

[2]Optimization problem is solvable if it is feasible, bounded below (for a minimization problem), and its optimal value is attained [12, page 15].

Unfortunately, there often is a nonzero gap between QMI problem (18) and its LMI relaxation problem (19). In that case, we wish either to use an optimal solution of the LMI problem to generate an approximate solution for the QMI problem within polynomial time complexity, or to find a sufficient optimality condition such that a globally optimal solution for the QMI problem is secured.

Evidently, if the matrix component $\boldsymbol{W}^\star = \boldsymbol{w}^\star\boldsymbol{w}^{\star H}$ of an optimal solution $(\boldsymbol{W}^\star, \boldsymbol{Y}^\star)$ for LMI relaxation problem (19) is of rank one, then one can claim that $\boldsymbol{w}^\star$ is also optimal for the QMI problem (18) (and thus for (12)). However, if the rank of $\boldsymbol{W}^\star$ is of rank two or above, we still have to find a suboptimal solution and characterize it.

To proceed, problem (19) is first transformed into the following equivalent problem:

$$\begin{array}{ll} \underset{\boldsymbol{W},\boldsymbol{Y},t}{\text{maximize}} & \text{tr}\,(\hat{\boldsymbol{R}}_s\boldsymbol{Y}) - \epsilon t \\ \text{subject to} & \text{tr}\left((\hat{\boldsymbol{R}} + \gamma\boldsymbol{I})\boldsymbol{W}\right) = 1 \\ & \boldsymbol{W} - \boldsymbol{Y} \succeq \boldsymbol{0} \\ & \|\boldsymbol{Y}\| \le t,\ \boldsymbol{W} \succeq \boldsymbol{0}. \end{array} \tag{20}$$

To compute the dual problem of problem (19), we claim the following lemma.

**Lemma III.2** *The dual problem of problem* (19) *is the following linear conic programming problem:*

$$\begin{array}{ll} \underset{z,\boldsymbol{Z}}{\text{minimize}} & z \\ \text{subject to} & \|\boldsymbol{Z} - \hat{\boldsymbol{R}}_s\| \le \epsilon, \\ & z\left(\hat{\boldsymbol{R}} + \gamma\boldsymbol{I}\right) - \boldsymbol{Z} \succeq \boldsymbol{0}, \\ & \boldsymbol{Z} \succeq \boldsymbol{0}. \end{array} \tag{21}$$

The lemma can be proved in a way similar to that of Lemma III.1 and thus we omit it for brevity.

Problems (20) and (21) are strictly feasible. For example, $\left(\boldsymbol{I}/\text{tr}\left((\hat{\boldsymbol{R}} + \gamma\boldsymbol{I})\right), \boldsymbol{0}, 1\right)$ and $\left(z, \hat{\boldsymbol{R}}_s + \epsilon\boldsymbol{I}/\sqrt{N+1}\right)$ with sufficiently large $z$ are strictly feasible points for (20) and (21), respectively. Thus, strong duality holds between (20) and (21).

It can be easily verified (see [12, Theorem 1.4.2]) that the complementary conditions (complementary slackness in the first-order optimality conditions, see [16]) for primal problem (20) and dual problem (21) are:

$$\epsilon t + \text{tr}\left(\boldsymbol{Y}(\boldsymbol{Z} - \hat{\boldsymbol{R}}_s)\right) = 0, \tag{22}$$

$$\text{tr}\left((z(\hat{\boldsymbol{R}} + \gamma\boldsymbol{I}) - \boldsymbol{Z})\boldsymbol{W}\right) = 0, \tag{23}$$

$$\text{tr}\,((\boldsymbol{W} - \boldsymbol{Y})\boldsymbol{Z}) = 0. \tag{24}$$

Observing further that $\text{tr}\,(\boldsymbol{W}(\hat{\boldsymbol{R}} + \gamma\boldsymbol{I})) = 1$, we arrive to the following compact form of the complementary conditions:

$$z = \text{tr}\,(\boldsymbol{W}\boldsymbol{Z}) = \text{tr}\,(\boldsymbol{Y}\boldsymbol{Z}) = \text{tr}\,(\boldsymbol{Y}\hat{\boldsymbol{R}}_s) - \epsilon\|\boldsymbol{Y}\|. \tag{25}$$

Note that the conditions in (25) are necessary and sufficient for a feasible primal-dual pair $\{\boldsymbol{W}, \boldsymbol{Y}; z, \boldsymbol{Z}\}$ to be optimal.

Suppose that $\{\boldsymbol{W}^\star, \boldsymbol{Y}^\star; z^\star, \boldsymbol{Z}^\star\}$ is an optimal primal-dual pair obtained, for example, by applying a primal-dual interior point method for solving (19) and (21). Suppose also that the rank of $\boldsymbol{W}^\star$ is greater than one. While looking for a rank-one approximate solution, we resort to the rank-one matrix decomposition lemma in [17], which is cited as follow.

**Lemma III.3 (Theorem 2.1 in [17])** *Suppose that $\boldsymbol{X}$ is a $N \times N$ complex Hermitian PSD matrix of rank $R$. Suppose also that $\boldsymbol{A}$ and $\boldsymbol{B}$ are two $N \times N$ given Hermitian matrices. Then, there exists a rank-one decomposition $\boldsymbol{X} = \sum_{r=1}^R \boldsymbol{x}_r \boldsymbol{x}_r^H$ such that*

$$\boldsymbol{x}_r^H \boldsymbol{A} \boldsymbol{x}_r = \frac{\text{tr}\,(\boldsymbol{AX})}{R} \quad \text{and} \quad \boldsymbol{x}_r^H \boldsymbol{B} \boldsymbol{x}_r = \frac{\text{tr}\,(\boldsymbol{BX})}{R}, \, \forall r$$

(*synthetically denoted as* $\mathcal{D}(\boldsymbol{X}, \boldsymbol{A}, \boldsymbol{B})$).

Leveraging Lemma III.3, we obtain $\boldsymbol{W}^\star = \sum_{i=1}^R \boldsymbol{w}_i \boldsymbol{w}_i^H$ (where $R$ is the rank of $\boldsymbol{W}^\star$) such that

$$\text{tr}\left((\hat{\boldsymbol{R}} + \gamma \boldsymbol{I})(R \boldsymbol{w}_i \boldsymbol{w}_i^H)\right) = \text{tr}\left((\hat{\boldsymbol{R}} + \gamma \boldsymbol{I}) \boldsymbol{W}^\star\right) = 1, \quad (26)$$

and

$$\text{tr}\left((R \boldsymbol{w}_i \boldsymbol{w}_i^H) \boldsymbol{Z}^\star\right) = \text{tr}\,(\boldsymbol{W}^\star \boldsymbol{Z}^\star) = z^\star, \, i = 1, \ldots, R. \quad (27)$$

Here (27) means that each $R \boldsymbol{w}_i \boldsymbol{w}_i^H$ (together with $\boldsymbol{Y}^\star$, $z^\star$, and $\boldsymbol{Z}^\star$) fulfills the optimality conditions stated in (25), while (26) implies that $\sqrt{R} \boldsymbol{w}_i$ complies only with the first constraint in (18). If some $\sqrt{R} \boldsymbol{w}_{i_0}$ (together with $\boldsymbol{Y}^\star$) satisfies the QMI constraint (the second constraint) in (18), then one can conclude that $\sqrt{R} \boldsymbol{w}_{i_0}$ is optimal for (18) due to (27).

In order to obtain the objective function value of (12) at $\sqrt{R} \boldsymbol{w}_i$, we substitute $R \boldsymbol{w}_i \boldsymbol{w}_i^H$ into (18) (which is equivalent to (12)), and get the following optimization problem:

$$\begin{array}{ll} \underset{\boldsymbol{Y}}{\text{maximize}} & \text{tr}\,(\hat{\boldsymbol{R}}_s \boldsymbol{Y}) - \epsilon \|\boldsymbol{Y}\| \\ \text{subject to} & R \boldsymbol{w}_i \boldsymbol{w}_i^H - \boldsymbol{Y} \succeq \boldsymbol{0}, \end{array} \quad (28)$$

where the first constraint in (18) is dropped since it is always satisfied thanks to (26).

Now we assume that $\bar{\boldsymbol{Y}}_i$ is an optimal solution for (28) associated with $\boldsymbol{w}_i$. If $\bar{\boldsymbol{Y}}_i$ fulfills the optimality conditions (25) (together with $R \boldsymbol{w}_i \boldsymbol{w}_i^H$, $\boldsymbol{Z}^\star$ and $z^\star$), then we conclude that $\sqrt{R} \boldsymbol{w}_i$ is optimal for (18).

Suppose that $\{\bar{\boldsymbol{Y}}_i\}_{i=1}^R$ are the optimal solutions for (28) corresponding to $\boldsymbol{w}_i$, $i = 1, \ldots, R$, respectively. Select

$$\bar{\boldsymbol{Y}}_{i_0} := \arg\max\{\text{tr}\,(\hat{\boldsymbol{R}}_s \bar{\boldsymbol{Y}}_i) - \epsilon \|\bar{\boldsymbol{Y}}_i\| \mid \bar{\boldsymbol{Y}}_i, \, i = 1, \ldots, R\},$$

together with the $R \boldsymbol{w}_{i_0} \boldsymbol{w}_{i_0}^H$. Thus, we return $(R \boldsymbol{w}_{i_0} \boldsymbol{w}_{i_0}^H, \bar{\boldsymbol{Y}}_{i_0})$ as a suboptimal solution for (19). In other words, $(\sqrt{R} \boldsymbol{w}_{i_0}, \bar{\boldsymbol{Y}}_{i_0})$ is feasible for (18), and $\sqrt{R} \boldsymbol{w}_{i_0}$ is an approximate solution for (12) (and thus also for original robust adaptive beamforming problem for general-rank signal model (6)).

Algorithm 1 summarizes the above described procedure of generating a suboptimal solution for (6).

The computational complexity of Algorithm 1 is dominated by solving SDP problem (20) and $R$ SDP problems (28), each of which has the worst-case complexity in $O(N^{6.5})$ (see [12]).

---

**Algorithm 1** Procedure for for finding a solution of robust beamforming problem for general-rank signal model (6).

---

**Input:** $\hat{\boldsymbol{R}}$, $\hat{\boldsymbol{R}}_s$, $\gamma$, $\epsilon$;
**Output:** A suboptimal solution $\boldsymbol{w}$ for (6);

1: solve SDP (20) and find the optimal solution $(\boldsymbol{W}^\star, \boldsymbol{Y}^\star)$ together with the dual optimal solution $(\boldsymbol{Z}^\star, z^\star)$;
2: if $\boldsymbol{W}^\star = \boldsymbol{w}\boldsymbol{w}^H$, i.e., $\boldsymbol{W}^\star$ has rank one, then output $\boldsymbol{w}$ as an optimal solution and terminate;
3: implement the rank-one decomposition $\mathcal{D}(\boldsymbol{W}^\star, \hat{\boldsymbol{R}} + \gamma \boldsymbol{I}, \boldsymbol{Z}^\star)$ and obtain $\boldsymbol{W}^\star = \sum_{i=1}^R \boldsymbol{w}_i \boldsymbol{w}_i^H$;
4: solve SDPs (28) and obtain $\boldsymbol{Y}_i^\star$, $i = 1, \ldots, R$;
5: compute $\boldsymbol{Y}_{i_0} := \arg\max\{\text{tr}\,(\hat{\boldsymbol{R}}_s \boldsymbol{Y}_i^\star) - \epsilon \|\boldsymbol{Y}_i^\star\| \mid \boldsymbol{Y}_i^\star, \, i = 1, \ldots, R\}$;
6: output $\sqrt{R} \boldsymbol{w}_{i_0}$.

---

## IV. Sufficient Optimality Conditions for QMI Problem (18)

Hereafter we assume that the eigenvalues of a Hermitian matrix are always placed in a descending order: $\lambda_{\max} = \lambda_1 \geq \lambda_2 \geq \ldots \geq \lambda_N = \lambda_{\min}$.

The following proposition is of importance.

**Lemma IV.1** *There holds:*

$$\left\{\boldsymbol{X} \in \mathcal{S}^N \mid \sqrt{N-1} \|\boldsymbol{X}\| \leq \text{tr}\,(\boldsymbol{X})\right\} \subseteq \mathcal{S}_+^N,$$

*where $\mathcal{S}_+^N$ is the cone of real symmetric PSD matrices. This also holds for $\mathcal{H}^N$ and $\mathcal{H}_+^N$.*

For completeness, we present a proof in Appendix B.

With Lemma IV.1 at hand, we can claim our main result of this section about a sufficient condition for the existence of rank-one solutions for problem (19), i.e., a sufficient optimality condition for QMI problem (18).

**Theorem IV.2** *Suppose that $(\boldsymbol{W}, \boldsymbol{Y})$ is optimal for problem* (19). *If*

$$\text{tr}\,(\boldsymbol{W} - \boldsymbol{Y}) \geq \text{tr}\,(\boldsymbol{W}) \sqrt{N-1} \left(1 + \frac{\lambda_1(\hat{\boldsymbol{R}}_s)}{\epsilon}\right). \quad (29)$$

*then there exists a rank-one solution for* (19).

See Appendix C for the proof.

The following corollary of the Theorem IV.2 is also instrumental in the paper.

**Corollary IV.3** *Suppose that $(\boldsymbol{W}, \boldsymbol{Y})$ is optimal for problem* (19). *If*

$$\text{tr}\,(\boldsymbol{Y}) \leq \frac{1}{\gamma + \lambda_1(\hat{\boldsymbol{R}})} - \frac{\sqrt{N-1}}{\gamma + \lambda_N(\hat{\boldsymbol{R}})} \left(1 + \frac{\lambda_1(\hat{\boldsymbol{R}}_s)}{\epsilon}\right), \quad (30)$$

*then there exists a rank-one solution for* (19).

See Appendix D for the proof.

We remark that the proof of Theorem IV.2 includes a solution procedure for obtaining a rank-one solution (via Lemma III.3) from a general-rank solution $\boldsymbol{W}$, where $(\boldsymbol{W}, \boldsymbol{Y})$ complies with (29). In other words, $\boldsymbol{w}\boldsymbol{w}^H$ is constructed (as in the proof) such that (46) and (47) are satisfied. Then it follows from Theorem IV.2 that if the sufficient condition (29) is fulfilled for $(\boldsymbol{W}, \boldsymbol{Y})$, then $(\boldsymbol{w}\boldsymbol{w}^H, \boldsymbol{Y})$ is optimal (rather than an approximate) solution for problem (19).

Condition (29) can be further refined as stated in the following corollary, if the optimal value is taken into consideration.

**Corollary IV.4** *Suppose that $(\boldsymbol{W}, \boldsymbol{Y})$ is optimal for problem* (19)*, and let $v^\star$ be the corresponding optimal value. If*

$$\operatorname{tr}(\boldsymbol{W} - \boldsymbol{Y}) \geq \operatorname{tr}(\boldsymbol{W})\sqrt{N-1}\left(1 + \frac{\lambda_1(\hat{\boldsymbol{R}}_s)}{\epsilon} - \frac{v^\star}{\epsilon \operatorname{tr}(\boldsymbol{W})}\right), \tag{31}$$

*then a rank-one solution for LMI relaxation problem* (19) *can be constructed.*

See Appendix E for the proof.

## V. A revisit of dual SDP problem (21)

In this section, we revisit dual problem (21) of LMI relaxation problem (19) for original QMI robust adaptive beamforming problem for general-rank signal model (18).

We start from the observation that the second constraint in (21) is equivalent to the following matrix inequality:

$$z\boldsymbol{I} \succeq \boldsymbol{Z}\left(\hat{\boldsymbol{R}} + \gamma \boldsymbol{I}\right)^{-1}.$$

Therefore, dual problem (21) can be interpreted in the way that the largest eigenvalue of $\boldsymbol{Z}(\hat{\boldsymbol{R}} + \gamma\boldsymbol{I})^{-1}$ is minimized over the set of $\boldsymbol{Z}$ defined by the other two constraints $\|\hat{\boldsymbol{R}}_s - \boldsymbol{Z}\| \leq \epsilon$ and $\boldsymbol{Z} \succeq \boldsymbol{0}$ in (21). Namely, (21) is equivalent to the following problem:

$$\begin{array}{ll} \underset{\boldsymbol{Z}}{\text{minimize}} & \lambda_1\left(\boldsymbol{Z}\left(\hat{\boldsymbol{R}} + \gamma\boldsymbol{I}\right)^{-1}\right) \\ \text{subject to} & \|\boldsymbol{Z} - \hat{\boldsymbol{R}}_s\| \leq \epsilon, \\ & \boldsymbol{Z} \succeq \boldsymbol{0}. \end{array} \tag{32}$$

It is known that

$$\lambda_1\left(\boldsymbol{Z}\left(\hat{\boldsymbol{R}} + \gamma\boldsymbol{I}\right)^{-1}\right) = \underset{\boldsymbol{w} \neq \boldsymbol{0}}{\text{maximize}} \ \frac{\boldsymbol{w}^H\boldsymbol{Z}\boldsymbol{w}}{\boldsymbol{w}^H\left(\hat{\boldsymbol{R}} + \gamma\boldsymbol{I}\right)\boldsymbol{w}}. \tag{33}$$

Let us set

$$\boldsymbol{\Delta}_2' = \boldsymbol{Z} - \hat{\boldsymbol{R}}_s. \tag{34}$$

To simplify the notation, we drop the superscript $'$ in $\boldsymbol{\Delta}_2'$ in the following discussion, without leading to confusions. Thus, (33) is recast into:

$$\underset{\boldsymbol{w} \neq \boldsymbol{0}}{\text{maximize}} \ \frac{\boldsymbol{w}^H\left(\hat{\boldsymbol{R}}_s + \boldsymbol{\Delta}_2\right)\boldsymbol{w}}{\boldsymbol{w}^H\left(\hat{\boldsymbol{R}} + \gamma\boldsymbol{I}\right)\boldsymbol{w}}. \tag{35}$$

Using (33), (34), and (35), problem (32) can be reexpressed as:

$$\underset{\boldsymbol{\Delta}_2 \in \mathcal{B}_2}{\text{minimize}} \ \underset{\boldsymbol{w} \neq \boldsymbol{0}}{\text{maximize}} \ \frac{\boldsymbol{w}^H\left(\hat{\boldsymbol{R}}_s + \boldsymbol{\Delta}_2\right)\boldsymbol{w}}{\boldsymbol{w}^H\left(\hat{\boldsymbol{R}} + \gamma\boldsymbol{I}\right)\boldsymbol{w}}, \tag{36}$$

where $\mathcal{B}_2$ has been defined in (8). Also, problem (36) can be further recast into:

$$\underset{\boldsymbol{\Delta}_2 \in \mathcal{B}_2}{\text{minimize}} \ \underset{\boldsymbol{w} \neq \boldsymbol{0}}{\text{maximize}} \ \underset{\boldsymbol{\Delta}_1 \in \mathcal{B}_1}{\text{minimize}} \ \frac{\boldsymbol{w}^H\left(\hat{\boldsymbol{R}}_s + \boldsymbol{\Delta}_2\right)\boldsymbol{w}}{\boldsymbol{w}^H\left(\hat{\boldsymbol{R}} + \boldsymbol{\Delta}_1\right)\boldsymbol{w}}, \tag{37}$$

In other words, dual SDP (21) is equivalent to (37). Therefore, we conclude the following proposition.

**Proposition V.1** *The SDP relaxation problem* (19) *is equivalent to the min-max-min problem* (37)*, in the sense that they share the same optimal value.*

Recall that problem (18) is a tantamount QMI formulation of the original robust beamforming (6), which is cited as follows.

$$\underset{\boldsymbol{w} \neq \boldsymbol{0}}{\text{maximize}} \ \underset{\boldsymbol{\Delta}_1 \in \mathcal{B}_1, \boldsymbol{\Delta}_2 \in \mathcal{B}_2}{\text{minimize}} \ \frac{\boldsymbol{w}^H\left(\hat{\boldsymbol{R}}_s + \boldsymbol{\Delta}_2\right)\boldsymbol{w}}{\boldsymbol{w}^H\left(\hat{\boldsymbol{R}} + \boldsymbol{\Delta}_1\right)\boldsymbol{w}} \tag{38}$$

Taking into consideration Proposition V.1, and the fact that (19) is the SDP relaxation for (18) (thus for (6)), we reach the following conclusion that is formulated also as a proposition.

**Proposition V.2** *The min-max-min problem* (37) *is an SDP relaxation for the maximin problem* (38).

It is known that another upper bound for (38) is the following minimax problem:

$$\underset{\boldsymbol{\Delta}_1 \in \mathcal{B}_1, \boldsymbol{\Delta}_2 \in \mathcal{B}_2}{\text{minimize}} \ \underset{\boldsymbol{w} \neq \boldsymbol{0}}{\text{maximize}} \ \frac{\boldsymbol{w}^H\left(\hat{\boldsymbol{R}}_s + \boldsymbol{\Delta}_2\right)\boldsymbol{w}}{\boldsymbol{w}^H\left(\hat{\boldsymbol{R}} + \boldsymbol{\Delta}_1\right)\boldsymbol{w}}. \tag{39}$$

It is easy to see that

$$v^\star((39)) \geq v^\star((37)) \geq v^\star((38)). \tag{40}$$

Observe also that $v^\star((37)) = v^\star((19))$ and $v^\star((38)) = v^\star((6))$. Then, we can claim a sandwich result as in the following theorem.

**Theorem V.3** *Minimax problem* (39) *is an upper bound for SDP* (19)*, which is an upper bound for maximin problem* (6).

We remark that Theorem V.3 indicates that the optimal value of original robust beamforming problem (6) has two upper bounds: the optimal value of the SDP relaxation (19) and the optimal value of (39). The SDP relaxation (19) is tighter than the minimax relaxation problem (39).

## VI. Simulation Results

We consider the scenario with a uniform linear array of $N = 10$ omnidirectional sensors spaced half a wavelength apart. The additive noise variance in each sensor is set to 0 dB. A signal from an interference source with the interference-to-noise ratio (INR) 30 dB impinges on the sensor array. Both the desired signal and the interference are locally incoherently scattered sources [18]. The signal of interest and the interference have Gaussian and uniform angular power densities with the central angles 30° and 10°, respectively, and the angular spreads 4° and $10°$, respectively. The presumed signal of interest is assumed to have Gaussian angular power density with central angle and angular spread 34° and $6°$, respectively.

The performance of the new proposed QMI beamforming method is compared to that of the beamformers proposed in [4]–[6], [9], termed hereafter as "New Beamformer", "SGLW Beamfomer", "CG Beamformer", "XMW Beamformer", and "KV Beamformer", respectively. All beanformers use the same sample data covariance matrix that is estimated with $T = 50$ snapshots. The diagonal loading parameter $\gamma = 0.1\|\hat{\boldsymbol{R}}\|$ is set and the norm bound $\epsilon = 0.3\|\hat{\boldsymbol{R}}_s\|$ is chosen herein. The norm bound $\eta = 0.9\sqrt{\mathrm{tr}\left(\hat{\boldsymbol{R}}_s\right)}$ is selected for the methods in [4]–[6], [9]. All results are averaged over 100 simulation trials.

*Example 1*: This example examines the beamformer output SINR versus SNR among the aforementioned beamformers tested. The simulation results are shown in Fig. 1 in the figure. It can be seen from the figure that the new QMI beamforming method leads to better performance than the ones in [4]–[6], [9]. It can be attributed to the fact that the new QMI robust beamforming methods solves the actual robust adaptive beamforming problem for general-rank signal model, while the competitive methods solve the modifications of the original problem.

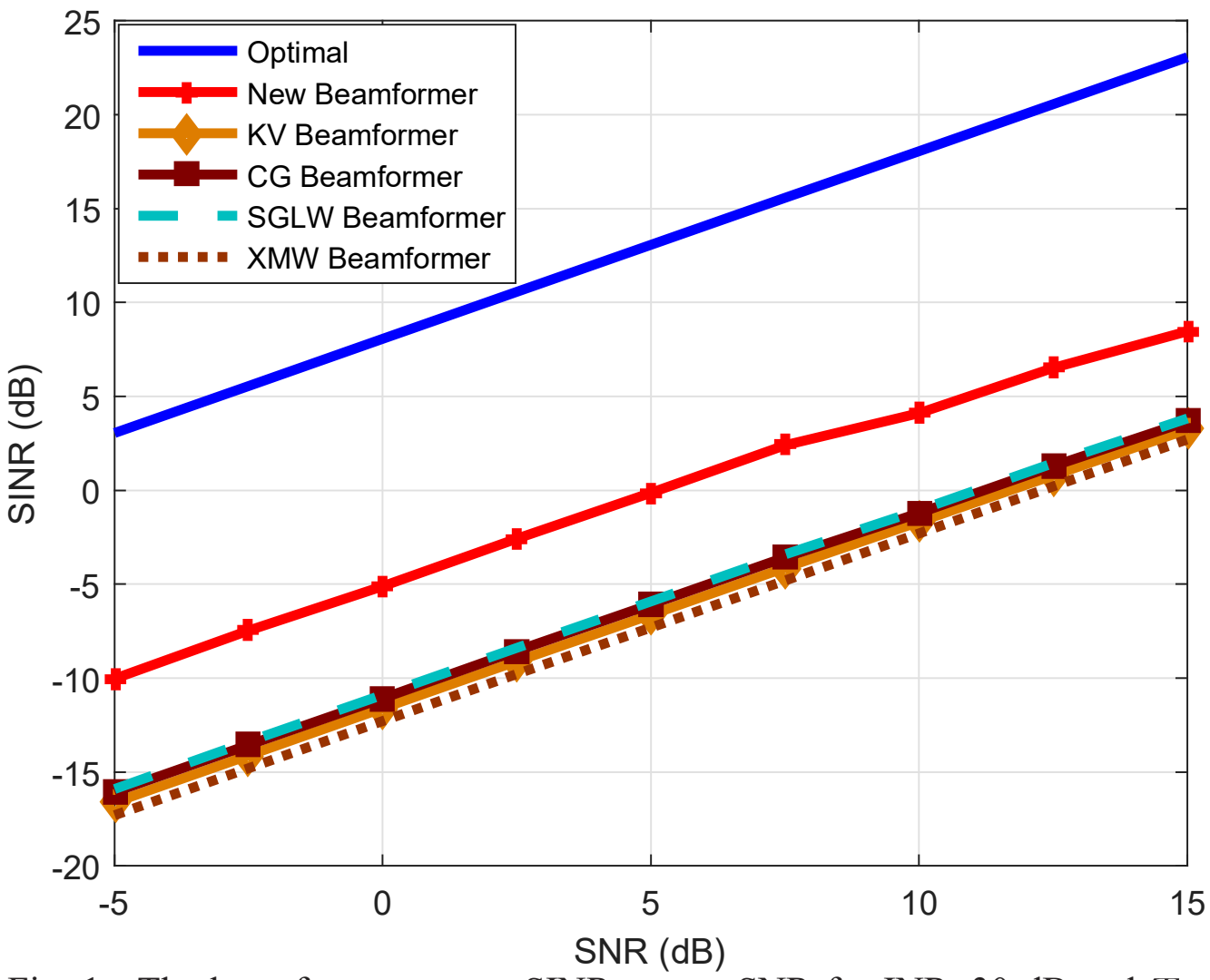

Fig. 1. The beamformer output SINR versus SNR for INR=30 dB and $T = 50$.

*Example 2*: It is known that when the angle spread of the desired source varies, the rank of the actual covariance $\boldsymbol{R}_s$ of the desired source changes, which can affect the performance of the beamformer. This example tests how much the output SINR can fluctuate if the angle spread is set to 0.15°, 1°, 2°, 5°, 9°, 14°, 20°, 25°, and 30° (the corresponding ranks of $\boldsymbol{R}_s$ are 2:1:10, respectively). Let SNR be set in this example as SNR=10 dB. All other simulation settings are the same as those explained in the beginning of this section. Fig. 2 plots the output SINRs versus the rank of the actual correlation matrix $\boldsymbol{R}_s$ for all aforementioned beamformers tested. It can be observed from the figure that the new QMI robust beamforming method again outperforms the ones proposed in [4]–[6], [9].

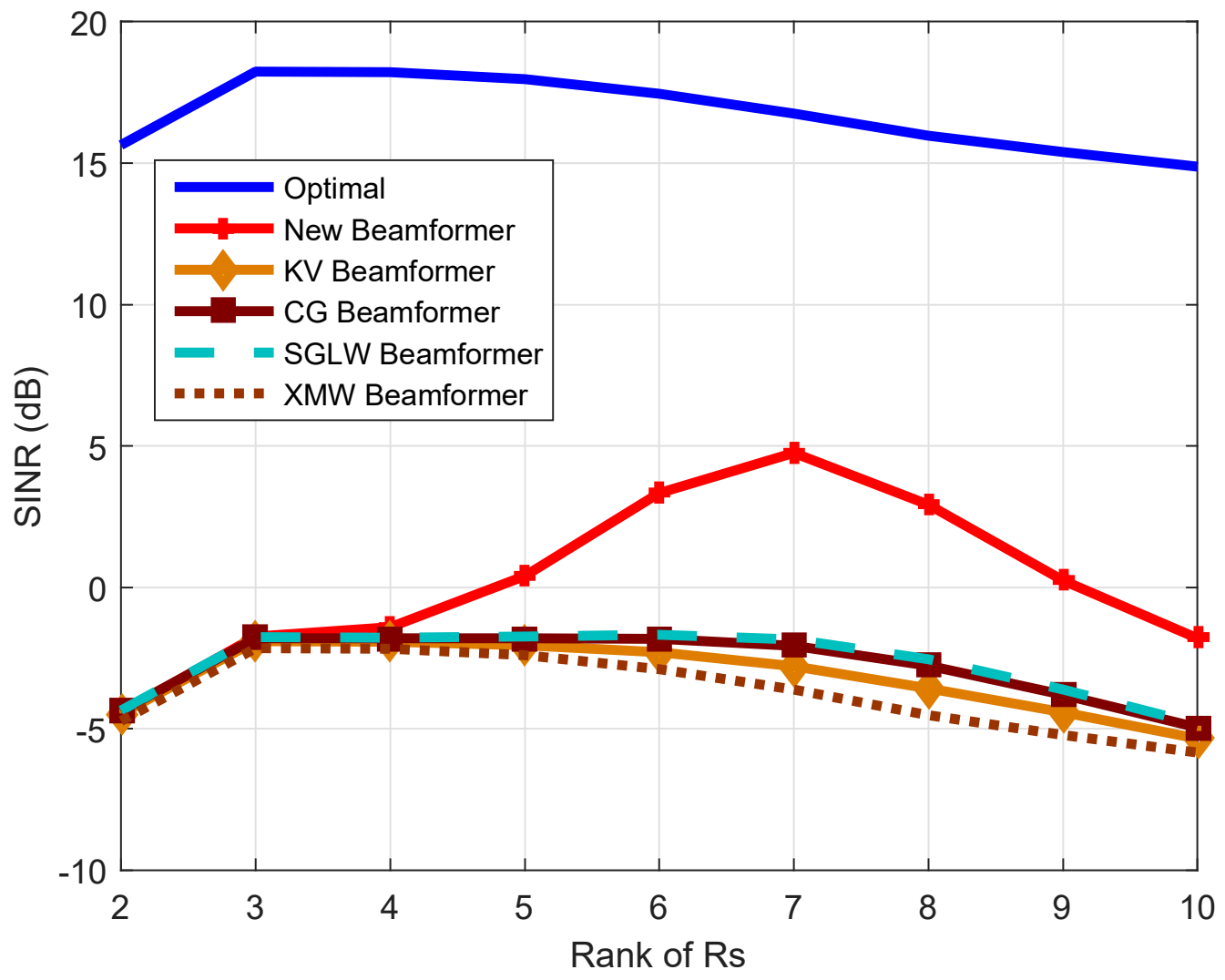

Fig. 2. The beamformer output SINR versus the rank of $\boldsymbol{R}_s$ for INR=30 dB, SNR=10 dB, and $T = 50$.

*Example 3*: In this example, we assume that the desired source angular power density is a truncated Laplacian function distorted by severe fluctuations, with the central angle and the scale parameter of the Laplacian distribution being $30°$ and $0.1$, respectively. The truncated probability density function of the distribution becomes zero outside interval $[15°, 45°]$. Both the presumed shape of the desired signal angular power density and the interference source are exactly modeled as those described in the beginning of this section. Fig. 3 depicts the output SINRs of the beamformer tested vesus SNR. This figure shows as well that the output SINR by the proposed QMI beamforming method is higher than those by the methods of [4]–[6], [9].

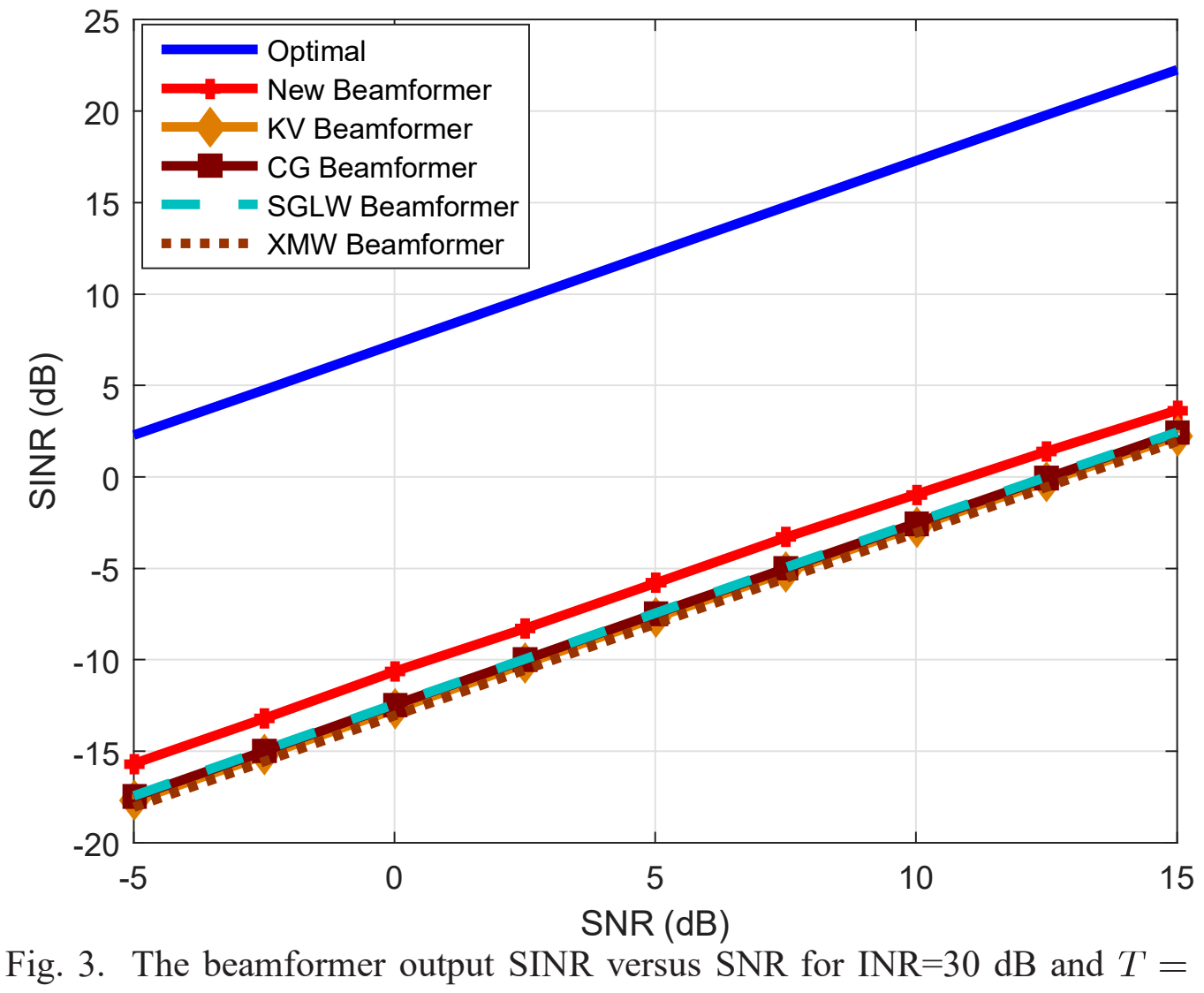

Fig. 3. The beamformer output SINR versus SNR for INR=30 dB and $T = 50$.

## VII. Conclusion

We have considered the robust adaptive beamforming problem for general-rank signal model. Unlike solving this problem by introducing a matrix decomposition on the presumed covariance matrix of the signal, as it has been previous done, we have studied the original robust adaptive beamforming problem. Resorting to the strong duality theorem for the linear conic programming, we have reformulated the beamforming problem into a nonconvex QMI problem, and relaxed it into a convex LMI problem. Due to the nonzero gap between the QMI and LMI problems, we have proposed an approximate algorithm with polynomial-time computational complexity in order to tackle the problem. In addition, sufficient optimality conditions for the nonconvex QMI problem have been established. The improved performance of the proposed robust adaptive beamformer for general-rank signal model has been demonstrated by simulations in terms of the beamformer output SINR.

Despite a specific QMI problem[3] has been addressed in the paper, the structure of the problem is quite typical in other applications of robust signal processing and optimization where the energy/variance needs to be minimized for the worst-case uncertainties in the second-order statistics about a general-rank signal subject to quadratic equality/inequality constraints. Thus, this problem is of a general interest in signal processing as well. The proposed QMI-based framework can be applied for solving other QMI problems.

## Appendix A
## Proof of Proposition II.1

Let $v_1$ and $v_2$ be the optimal values for (11) and (12), respectively. Reexpress also (12) equivalently as

$$\begin{array}{ll}\underset{\boldsymbol{w}\neq\boldsymbol{0}}{\text{maximize}} & \dfrac{\underset{\boldsymbol{\Delta}_2\in\mathcal{B}_2}{\text{minimize}}\;\boldsymbol{w}^H(\hat{\boldsymbol{R}}_s+\boldsymbol{\Delta}_2)\boldsymbol{w}}{\boldsymbol{w}^H(\hat{\boldsymbol{R}}+\gamma\boldsymbol{I})\boldsymbol{w}} \\ \text{subject to} & \boldsymbol{w}^H(\hat{\boldsymbol{R}}+\gamma\boldsymbol{I})\boldsymbol{w}=1.\end{array} \tag{41}$$

Therefore, it can be seen that

$$v_1 \geq v_2, \tag{42}$$

since the feasible set of (41) is smaller than that of (11).

Let us define the following function

$$f(\boldsymbol{w}) = \frac{\underset{\boldsymbol{\Delta}_2\in\mathcal{B}_2}{\text{minimize}}\;\boldsymbol{w}^H(\hat{\boldsymbol{R}}_s+\boldsymbol{\Delta}_2)\boldsymbol{w}}{\boldsymbol{w}^H(\hat{\boldsymbol{R}}+\gamma\boldsymbol{I})\boldsymbol{w}}.$$

Suppose that $\boldsymbol{w}^\star$ is an optimal solution for (11). Define

$$\boldsymbol{u}^\star = \frac{\boldsymbol{w}^\star}{\|(\hat{\boldsymbol{R}}+\gamma\boldsymbol{I})^{1/2}\boldsymbol{w}^\star\|}.$$

It follows that

$$f(\boldsymbol{u}^\star) = f(\boldsymbol{w}^\star) = v_1.$$

In other words, $\boldsymbol{u}^\star$ is optimal for (11), with $\boldsymbol{u}^{\star H}(\hat{\boldsymbol{R}}+\gamma\boldsymbol{I})\boldsymbol{u}^\star = 1$. Therefore, $\boldsymbol{u}^\star$ is feasible for (41) (i.e. (12)), and then we have

$$v_2 \geq f(\boldsymbol{u}^\star) = v_1. \tag{43}$$

By (42) and (43), we conclude that $v_1 = v_2$.

[3]It is the original robust adaptive beamforming problem for general rank signal model (12) (or (11)).

Since problems (11) and (41) have the same objective function, and $v_1 = v_2$, hence an optimal solution $\boldsymbol{w}^\star$ for (41) (i.e. (12)) is also optimal for (11). The proof is thus complete. □

## Appendix B
## Proof of Lemma IV.1

We show the inclusion:

$$\mathcal{A} = \left\{\boldsymbol{X}\in\mathcal{S}^N | \sqrt{N-1}\,\|\boldsymbol{X}\| \leq \operatorname{tr}(\boldsymbol{X})\right\} \subseteq \mathcal{S}_+^N.$$

Suppose $\boldsymbol{X}\in\mathcal{A}$ and its eigenvalues are $\lambda_1 \geq \cdots \geq \lambda_N$. Thus, the condition $\sqrt{N-1}\|\boldsymbol{X}\| \leq \operatorname{tr}(\boldsymbol{X})$ implies that

$$\sqrt{N-1}\sqrt{\lambda_1^2+\cdots+\lambda_N^2} \leq \lambda_1+\cdots+\lambda_N. \tag{44}$$

We wish to show that $\lambda_1 \geq 0, \ldots, \lambda_N \geq 0$, given (44).

It can be seen that when $N = 1$, we obtained $\lambda_1 \geq 0$. Assume now that $N \geq 2$, and suppose that there exists at least one negative eigenvalue, thst is,

$$\lambda_1 \geq \cdots \geq \lambda_K \geq 0 > \lambda_{K+1} \geq \cdots \geq \lambda_N$$

for some $K$ satisfying $1 \leq K \leq N-1$ (if $K = N$, then all eigenvalues are nonnegative and the proof is done).

It follows from (44) that

$$\begin{aligned}\frac{\left(\sum_{n=1}^N \lambda_n\right)^2}{N-1} \geq \sum_{n=1}^N \lambda_n^2 &= \sum_{n=1}^K \lambda_n^2 + \sum_{n=K+1}^N \lambda_n^2 \\ &\geq \frac{\left(\sum_{n=1}^K \lambda_n\right)^2}{K} + \sum_{n=K+1}^N \lambda_n^2\end{aligned}$$

Therefore,

$$\frac{\left(\sum_{n=1}^N \lambda_n\right)^2}{N-1} - \frac{\left(\sum_{n=1}^K \lambda_n\right)^2}{K} \geq \sum_{n=K+1}^N \lambda_n^2$$

Since $1 \leq K \leq N-1$, hence

$$\frac{\left(\sum_{n=1}^N \lambda_n\right)^2}{N-1} - \frac{\left(\sum_{n=1}^K \lambda_n\right)^2}{N-1} \geq \sum_{n=K+1}^N \lambda_n^2$$

Then by

$$\begin{aligned}\left(\sum_{n=1}^N \lambda_n\right)^2 = \left(\sum_{n=1}^K \lambda_n\right)^2 &+ 2\left(\sum_{n=1}^K \lambda_n\right)\left(\sum_{n=K+1}^N \lambda_n\right) \\ &+ \left(\sum_{n=K+1}^N \lambda_n\right)^2\end{aligned}$$

we have

$$\begin{aligned}&\frac{\left(\sum_{n=1}^N \lambda_n\right)^2 - \left(\sum_{n=1}^K \lambda_n\right)^2}{N-1} \\ &= \frac{2\left(\sum_{n=1}^K \lambda_n\right)\left(\sum_{n=K+1}^N \lambda_n\right) + \left(\sum_{n=K+1}^N \lambda_n\right)^2}{N-1} \\ &\geq \sum_{n=K+1}^N \lambda_n^2\end{aligned}$$

Observe that

$$(N-K)\sum_{n=K+1}^{N}\lambda_n^2 \geq \left(\sum_{n=K+1}^{N}\lambda_n\right)^2.$$

Thus

$$\begin{aligned} 2\left(\sum_{n=1}^{K}\lambda_n\right)\left(\sum_{n=K+1}^{N}\lambda_n\right) &+ (N-K)\sum_{n=K+1}^{N}\lambda_n^2 \\ &\geq (N-1)\sum_{n=K+1}^{N}\lambda_n^2, \end{aligned}$$

which means

$$2\left(\sum_{n=1}^{K}\lambda_n\right)\left(\sum_{n=K+1}^{N}\lambda_n\right) \geq (K-1)\sum_{n=K+1}^{N}\lambda_n^2. \qquad (45)$$

We proceed now by discussing of the following three possible cases.

Case 1. Assume that $\lambda_K > 0 > \lambda_{K+1}$. It can be easily verified that in (45), the left-hand side is negative but the right-hand is nonnegative. Clearly it is a contradiction.

Case 2. Assume that $\lambda_1 \geq \cdots \geq \lambda_L > 0 = \lambda_{L+1} = \cdots = \lambda_K > \lambda_{K+1}$ for $L \geq 2$. Again the same contradiction can be deduced.

Case 3. Assume that $0 = \lambda_1 = \cdots = \lambda_K > \lambda_{K+1}$. It follows from (44) that

$$\sqrt{N-1}\sqrt{\lambda_{K+1}^2+\cdots+\lambda_N^2} \leq \lambda_{K+1}+\cdots+\lambda_N,$$

which leads to a contradiction too.

Therefore, we conclude that all eigenvalues $\lambda_1, \ldots, \lambda_N$, are nonnegative; namely, $\boldsymbol{X} \in \mathcal{S}_+^N$. The proof is complete. □

## Appendix C
## Proof of Theorem IV.2

We claim that $\text{tr}\,(\boldsymbol{W}) > 0$. In fact, if the trace is zero, then $\boldsymbol{W} = \boldsymbol{0}$, and $\text{tr}\,((\hat{\boldsymbol{R}}+\gamma\boldsymbol{I})\boldsymbol{W}) = 0 \neq 1$, which means that $\boldsymbol{W}$ is not feasible. But this is a contradiction because $(\boldsymbol{W},\boldsymbol{Y})$ is an optimal solution.

It follows from rank-one matrix decomposition Lemma III.3 that $\boldsymbol{W}$ there is a vector $\boldsymbol{w}$ such that

$$\boldsymbol{w}^H\boldsymbol{w} = \text{tr}\,(\boldsymbol{W}), \qquad (46)$$

and

$$\boldsymbol{w}^H(\hat{\boldsymbol{R}}+\gamma\boldsymbol{I})\boldsymbol{w} = \text{tr}\,\left((\hat{\boldsymbol{R}}+\gamma\boldsymbol{I})\boldsymbol{W}\right) = 1 \qquad (47)$$

(similar to (26) and (27)). Condition (29) together with equation (46) yields

$$\begin{aligned} \sqrt{N-1}\text{tr}\,(\boldsymbol{w}\boldsymbol{w}^H) &+ \sqrt{N-1}\frac{\lambda_1(\hat{\boldsymbol{R}}_s)\text{tr}\,(\boldsymbol{w}\boldsymbol{w}^H)}{\epsilon} \\ &\leq \text{tr}\,(\boldsymbol{w}\boldsymbol{w}^H) - \text{tr}\,(\boldsymbol{Y}). \end{aligned} \qquad (48)$$

Since $(\boldsymbol{W},\boldsymbol{Y})$ is an optimal solution, we have

$$\boldsymbol{W}-\boldsymbol{Y} \succeq \boldsymbol{0},$$

which implies that

$$\text{tr}\,(\boldsymbol{W}\hat{\boldsymbol{R}}_s) \geq \text{tr}\,(\boldsymbol{Y}\hat{\boldsymbol{R}}_s). \qquad (49)$$

Moreover, because we are maximizing the SINR, the optimal value has to be positive, i.e., $\text{tr}\,(\hat{\boldsymbol{R}}_s\boldsymbol{Y}) - \epsilon\|\boldsymbol{Y}\| > 0$, which implies that

$$\frac{\text{tr}\,(\hat{\boldsymbol{R}}_s\boldsymbol{Y})}{\|\boldsymbol{Y}\|} > \epsilon.$$

Then it follows that

$$\begin{aligned} \epsilon &< \frac{\text{tr}\,(\hat{\boldsymbol{R}}_s\boldsymbol{Y})}{\|\boldsymbol{Y}\|} \leq \frac{\text{tr}\,(\hat{\boldsymbol{R}}_s\boldsymbol{W})}{\text{tr}\,(\boldsymbol{W})}\frac{\text{tr}\,(\boldsymbol{W})}{\|\boldsymbol{Y}\|} \\ &\leq \lambda_1(\hat{\boldsymbol{R}}_s)\frac{\text{tr}\,(\boldsymbol{W})}{\|\boldsymbol{Y}\|} = \lambda_1(\hat{\boldsymbol{R}}_s)\frac{\text{tr}\,(\boldsymbol{w}\boldsymbol{w}^H)}{\|\boldsymbol{Y}\|}, \end{aligned}$$

where the second inequality holds true thanks to (49) and the fact that $\text{tr}\,(\boldsymbol{W}) \neq 0$. In other words, we have

$$\|\boldsymbol{Y}\| < \frac{\lambda_1(\hat{\boldsymbol{R}}_s)\text{tr}\,(\boldsymbol{w}\boldsymbol{w}^H)}{\epsilon}. \qquad (50)$$

Therefore, (48) and (50) give rise to

$$\sqrt{N-1}\text{tr}\,(\boldsymbol{w}\boldsymbol{w}^H) + \sqrt{N-1}\|\boldsymbol{Y}\| \leq \text{tr}\,(\boldsymbol{w}\boldsymbol{w}^H) - \text{tr}\,(\boldsymbol{Y}).$$

Using the fact that $\text{tr}\,(\boldsymbol{w}\boldsymbol{w}^H) = \|\boldsymbol{w}\boldsymbol{w}^H\|$, we further have

$$\sqrt{N-1}\|\boldsymbol{w}\boldsymbol{w}^H\| + \sqrt{N-1}\|\boldsymbol{Y}\| \leq \text{tr}\,(\boldsymbol{w}\boldsymbol{w}^H) - \text{tr}\,(\boldsymbol{Y}),$$

which implies that

$$\sqrt{N-1}\|\boldsymbol{w}\boldsymbol{w}^H - \boldsymbol{Y}\| \leq \text{tr}\,(\boldsymbol{w}\boldsymbol{w}^H) - \text{tr}\,(\boldsymbol{Y}).$$

It follows from Lemma IV.1 that

$$\boldsymbol{w}\boldsymbol{w}^H - \boldsymbol{Y} \succeq \boldsymbol{0} \qquad (51)$$

This, together with (47), implies that $(\boldsymbol{w}\boldsymbol{w}^H,\boldsymbol{Y})$ is feasible for problem (19), with the objective function value equal to the optimal value $\text{tr}\,(\hat{\boldsymbol{R}}_s\boldsymbol{Y}) - \epsilon\|\boldsymbol{Y}\|$. Therefore, we claim that $(\boldsymbol{w}\boldsymbol{w}^H,\boldsymbol{Y})$ is not only feasible, but also optimal for (19). The proof is thus complete. □

## Appendix D
## Proof of Corollary IV.3

It is known that $\text{tr}\,(\boldsymbol{W}) > 0$ (see the beginning of the proof of Theorem IV.2). Based on the fact that $\text{tr}\,((\hat{\boldsymbol{R}}+\gamma\boldsymbol{I})\boldsymbol{W}) = 1$, it follows that

$$\gamma = \frac{1}{\text{tr}\,(\boldsymbol{W})} - \frac{\text{tr}\,(\hat{\boldsymbol{R}}\boldsymbol{W})}{\text{tr}\,(\boldsymbol{W})} \leq \frac{1}{\text{tr}\,(\boldsymbol{W})} - \lambda_N(\hat{\boldsymbol{R}}),$$

which implies that

$$\text{tr}\,(\boldsymbol{W}) \leq \frac{1}{\gamma+\lambda_N(\hat{\boldsymbol{R}})}.$$

Similarly, we have

$$\text{tr}\,(\boldsymbol{W}) \geq \frac{1}{\gamma+\lambda_1(\hat{\boldsymbol{R}})}. \qquad (52)$$

Using condition (30), we obtain

$$\text{tr}\,(\boldsymbol{Y}) \leq \text{tr}\,(\boldsymbol{W}) - \text{tr}\,(\boldsymbol{W})\sqrt{N-1}\left(1+\frac{\lambda_1(\hat{\boldsymbol{R}}_s)}{\epsilon}\right), \qquad (53)$$

which amounts to

$$\text{tr}\,(\boldsymbol{W}-\boldsymbol{Y}) \geq \text{tr}\,(\boldsymbol{W})\sqrt{N-1}\left(1+\frac{\lambda_1(\hat{\boldsymbol{R}}_s)}{\epsilon}\right). \qquad (54)$$

It thus follows from Theorem IV.2 that there is a rank-one solution for (19) that completes the proof. □

## Appendix E
## Proof of Corollary IV.4

It follows that

$$\begin{aligned}\mathrm{tr}\,(\boldsymbol{W})\left(\frac{\lambda_1(\hat{\boldsymbol{R}}_s)}{\epsilon}-\frac{v^\star}{\epsilon \mathrm{tr}\,(\boldsymbol{W})}\right) &= \frac{\mathrm{tr}\,(\boldsymbol{W})\lambda_1(\hat{\boldsymbol{R}}_s)}{\epsilon}-\frac{v^\star}{\epsilon}\\ &\geq \frac{\mathrm{tr}\,(\hat{\boldsymbol{R}}_s\boldsymbol{W})-v^\star}{\epsilon}\\ &\geq \frac{\mathrm{tr}\,(\hat{\boldsymbol{R}}_s\boldsymbol{Y})-v^\star}{\epsilon}\\ &= \|\boldsymbol{Y}\|.\end{aligned}$$

Condition (31) implies that

$$\mathrm{tr}\,(\boldsymbol{W}-\boldsymbol{Y}) \geq \mathrm{tr}\,(\boldsymbol{W})\sqrt{N-1}+\sqrt{N-1}\|\boldsymbol{Y}\|. \tag{55}$$

The rest of the proof is similar to the proof of Theorem IV.2. For completeness, we present it in a fast manner as follows.

Using the fact that there is a rank-one matrix decomposition $\boldsymbol{W}=\sum_{i=1}^{R}\boldsymbol{w}_i\boldsymbol{w}_i^H$ (cf. Lemma III.3) such that $\mathrm{tr}\left(\sqrt{R}\boldsymbol{w}_i\sqrt{R}\boldsymbol{w}_i^H\right)=\mathrm{tr}\,(\boldsymbol{W})$, and

$$\left(\sqrt{R}\boldsymbol{w}_i\right)^H\left(\hat{\boldsymbol{R}}+\gamma\boldsymbol{I}\right)\left(\sqrt{R}\boldsymbol{w}_i\right)=\mathrm{tr}\left((\hat{\boldsymbol{R}}+\gamma\boldsymbol{I})\boldsymbol{W}\right)=1 \tag{56}$$

for all $i$, hence (55) implies that

$$\begin{aligned}\mathrm{tr}\,(\boldsymbol{w}\boldsymbol{w}-\boldsymbol{Y}) &\geq \mathrm{tr}\,(\boldsymbol{w}\boldsymbol{w}^H)\sqrt{N-1}+\sqrt{N-1}\|\boldsymbol{Y}\|\\ &= \sqrt{N-1}\left\|\boldsymbol{w}\boldsymbol{w}^H\right\|+\sqrt{N-1}\|\boldsymbol{Y}\|\\ &\geq \sqrt{N-1}\left\|\boldsymbol{w}\boldsymbol{w}^H-\boldsymbol{Y}\right\|\end{aligned}$$

for some $\boldsymbol{w}\in\left\{\sqrt{R}\boldsymbol{w}_i\right\}$. It therefore follows from Lemma IV.1 that

$$\boldsymbol{w}\boldsymbol{w}^H-\boldsymbol{Y}\succeq\boldsymbol{0}.$$

This, together with (56), implies that $(\boldsymbol{w}\boldsymbol{w}^H,\boldsymbol{Y})$ is optimal for (19) that completes the proof. □